\documentclass[a4paper,twocolumn,superscriptaddress,11pt,accepted=2019-02-20]{quantumarticle}
\usepackage[numbers,sort&compress]{natbib}
\usepackage[utf8]{inputenc}
\usepackage[english]{babel}
\usepackage[T1]{fontenc}
\usepackage{amsmath}
\usepackage{color}
\usepackage{graphicx}% Include figure files
\usepackage{dcolumn}% Align table columns on decimal point
\usepackage{array}
\usepackage{float}
\usepackage{longtable}
\usepackage{mathrsfs}
\usepackage{txfonts}
\usepackage{wasysym}
\usepackage[T1]{fontenc}
\usepackage{amssymb}
\usepackage[usenames,dvipsnames]{xcolor}
\usepackage{amsmath}
\usepackage{amstext}
\usepackage{latexsym}
\usepackage{hyperref}
\usepackage{bbold}

\usepackage{amsbsy} % for bold symbol 

\graphicspath{{../images/}{../gnuplot/}}

\begin{document}
\title{Work Statistics, Loschmidt Echo  and Information Scrambling in Chaotic Quantum Systems}

\author{Aur\'elia Chenu}
\email{achenu@dipc.org}
\homepage{http://dipc.ehu.es/chenu}
\orcid{0000-0002-4461-8289}
\affiliation{Donostia International Physics Center,  E-20018 San Sebasti\'an, Spain}
\affiliation{IKERBASQUE, Basque Foundation for Science, E-48013 Bilbao, Spain}
\affiliation{Theory Division, Los Alamos National Laboratory, MS-B213, Los Alamos, NM 87545, USA}
\affiliation{Massachusetts Institute of Technology, 77 Massachusetts Avenue, Cambridge, MA 02139, USA}
\author{Javier Molina-Vilaplana}
\orcid{0000-0002-9333-0062}
\affiliation{Technical University of Cartagena, UPCT, 30202, Cartagena, Spain}
\author{Adolfo del Campo}
\affiliation{Donostia International Physics Center,  E-20018 San Sebasti\'an, Spain}
\affiliation{IKERBASQUE, Basque Foundation for Science, E-48013 Bilbao, Spain}
\affiliation{Theory Division, Los Alamos National Laboratory, MS-B213, Los Alamos, NM 87545, USA}
\affiliation{Department of Physics, University of Massachusetts, Boston, MA 02125, USA}
%\date{\today}

\def\q{{\bf q}}

\def\G{\Gamma}
\def\L{\Lambda}
\def\la{\lambda}
\def\g{\gamma}
\def\al{\alpha}
\def\s{\sigma}
\def\e{\epsilon}
\def\k{\kappa}
\def\ve{\varepsilon}
\def\l{\left}
\def\r{\right}
\def\te{\mbox{e}}
\def\d{{\rm d}}
\def\t{{\rm t}}
\def\K{{\rm K}}
\def\N{{\rm N}}
\def\H{{\rm H}}
\def\la{\langle}
\def\ra{\rangle}
\def\om{\omega}
\def\Om{\Omega}
\def\vep{\varepsilon}
\def\wh{\widehat}
\def\tr{{\rm Tr}}
\def\da{\dagger}
\def\iz{\left}
\def\zi{\right}
\def\be{\boldsymbol{E}}
\def\bepsilon{\boldsymbol{\epsilon}}
\def\btheta{\boldsymbol{\theta}}
\newcommand{\beq}{\begin{equation}}
\newcommand{\eeq}{\end{equation}}
\newcommand{\beqa}{\begin{eqnarray}}
\newcommand{\eeqa}{\end{eqnarray}}
\newcommand{\intf}{\int_{-\infty}^\infty}
\newcommand{\into}{\int_0^\infty}

\newcommand{\ket}[1] {\vert #1 \rangle}
\newcommand{\bra}[1] {\langle #1 |}
\newcommand{\braket}[2] {\langle #1 | #2 \rangle}
\newcommand{\Tr}{\mathrm{Tr}}
\newcommand{\mix}{\mathrm{mix}}
\newcommand{\Th}{\mathrm{th}}

\newcommand{\aurelia}{\color{blue}}

\begin{abstract}

Characterizing the work statistics of driven complex quantum systems is generally challenging because of the exponential growth with the system size of the number of transitions involved between different energy levels. We consider the quantum work distribution associated with the driving of chaotic quantum systems described by random matrix Hamiltonians and characterize exactly the work statistics associated with a sudden quench for  arbitrary temperature and system size. 
Knowledge of the work statistics yields the Loschmidt echo dynamics of an entangled state between two copies of the system of interest, the thermofield double state. This echo dynamics  is dictated by the spectral form factor. We discuss its  relation to  frame potentials and its use to assess information scrambling.

\end{abstract}

\maketitle

Quantum thermodynamics has become a research field in bloom,  building on the dialogue between technological progress and foundations of physics \cite{Vinjanampathy2016a,Goold2016a}. In the presence of thermal and quantum fluctuations, familiar concepts from traditional thermodynamics at the macroscale,  such as heat and work,  become stochastic variables. Their analysis has been a fertile ground of inquiry, leading to the discovery of fluctuation theorems \cite{Jarzynski1997a,Crooks1999a,Tasaki2000a,Kurchan2000a,Campisi2011a} and time-work uncertainty relations \cite{Xiao2015a,Funo2017a,Bravetti2017a}.

In the quantum domain, even for isolated systems governed by unitary dynamics, work is not considered to be an observable. 
The definition of work requires  two projective measurements, one at the beginning and another at the end of the physical process under consideration \cite{Kurchan2000a,Talkner2007a}. Measuring the work statistics done by or on a driven system requires analyzing the transitions between its energy levels at the beginning and at the end of the process. In simple quantum systems,  this is an amenable task and pioneering experiments have measured the quantum work statistics, e.g., in coupled qubits in a NMR setting \cite{Batalhao2014a}, a driven quantum oscillator describing the dynamics of a trapped ion \cite{An2015a}, and ultracold atoms \cite{Cerisola2017a}.

For complex quantum systems, determining the work   probability density function  associated with a given process is intrinsically challenging, due to the exponential growth in the number of transitions with the dimension of the Hilbert space. 
Progress in this direction  has been achieved avoiding their explicit computation. To this end, it has proved useful to recast the characteristic function of the work distribution function as a Loschmidt echo amplitude \cite{Silva2008a}. 
This interpretation has facilitated the study of quantum work statistics in driven many-body spin systems  \cite{Silva2008a,Dorner2013a,Campbell2016a} and quantum fluids \cite{Garcia-March2016a}. 

Among complex systems, those exhibiting quantum chaos constitute a paradigmatic reference example. Following pioneering work by Wigner and Dyson, quantum chaotic systems are often defined as systems whose spectral properties are well described by random-matrix Hamiltonian ensembles \cite{MehtaBook,ForresterBook}. Random-matrix Hamiltonians can actually characterize a variety of physical systems ranging from atomic nuclei to spin chains. Quantum chaos plays as well a prominent role in black hole physics,  as it is widely believed that quantum correlations in these systems are scrambled at a maximum rate saturating an analogue of the Lyapunov exponent \cite{Maldacena2016a}.  It should be noted that, very recently, preliminary steps have been taken  to study work statistics in quantum chaotic systems \cite{Garcia-Mata2017a,Lobejko2017a,Chenu2018a}. 

In addition, black hole physics has motivated recent progress to characterize  information scrambling, i.e., the spreading of initially localized quantum information across different degrees of freedom in many-body systems.  Approaches to this end include the study of the decay dynamics of certain entangled states (the thermofield double state, associated with the purification of a thermal density matrix) \cite{Papadodimas2015a,Miyaji2016a,delCampo2017a},  related analytically-continued partition functions \cite{Cotler2017b, Dyer2017a},  out-of-time-order correlators \cite{Maldacena2016a} and frame potentials \cite{Cotler2017a}, some of which have been related to work statistics \cite{Chenu2018a,Campisi2017a,YungerHalpern2017a}. 

%Further, work on blackhole physics has led to the identification of complementary approaches to assess information scrambling, exploring the dynamics of e.g. entangled states , among other approaches. a mapping has been established between the problem of information scrambling and  work statistics \cite{Chenu17} as well as between work statistics and out-of-time order correlators.

In this work, we characterize the work statistics when a quantum chaotic system is driven  between two different chaotic Hamiltonians.  We show that the characteristic function  of the work probability distribution of a thermal state is equivalent to the Loschmidt echo of an entangled state.  Using this relation, we show that the work statistics can be used to elucidate the dynamics responsible for information scrambling in chaotic quantum systems. The connection between work statistics and  information scrambling is further discussed by  relating  the Loschmidt echo to the spectral form factor and frame potentials. 
%Our results are  nonperturbative and exact for Hamiltonians of any finite dimension sampled from the Gaussian Unitary Ensemble (GUE). In addition, we relate the characteristic function  of the work probability distribution  associated with the driving of a thermal state to the Loschmidt echo of a (purified) thermofield double state, that is characterized exactly.
% that undergoes a sudden quench between the initial and final Hamiltonians of the driving protocol. 

\section{Quantum work statistics in chaotic quantum systems}

\emph{Work probability distribution function.---} 
Let us consider  a driven, isolated  system described by the instantaneous Hamiltonian at time $s$ 
\beqa
\hat{H}_s=\sum_nE_n^s|n_s\ra\la n_s|\, ,
\eeqa
where $E_n^s$ denotes the real energy eigenvalue for the eigenstate $|n_s\ra$.
We focus on the time-evolution of a system initially prepared in a mixed state $\hat{\rho}$ at  time $s=0$ and driven to a final Hamiltonian $\hat{H}_\tau$ at time $s=\tau$. The corresponding time-evolution operator is 
\beqa
\label{Utau}
\hat{U}(\tau)={\cal T}\exp\left[-i\int_{0}^{\tau}ds \hat{H}_s\right]\, , 
\eeqa
where  the time-ordering operator ${\cal T}$ is required whenever the instantaneous Hamiltonians at two different times of evolution $0\leq (s,s') \leq\tau$ do not commute.
The associated work probability distribution function (work pdf, hereafter) is given by  \cite{Kurchan2000a,Talkner2007a}
\begin{equation}\label{defwork}
p_\tau(W):=\sum_{n,m} p^0_{n}\;  p^\tau_{m \vert n} \delta\left[W-\left(E_m^\tau -E_n^0 \right)\right]\, .
\end{equation}
Here, a first projective energy measurement leads to the probability $p^0_{n}= \langle n_0|\hat{\rho}|n_0\rangle$   to find the initial state $\hat{\rho}$ in the $n$-th eigenmode of the initial Hamiltonian $\hat{H}_0$. 
Following the evolution associated with $\hat{U}(\tau)$, a second projective energy measurement is used to determine the   transition probability from $|n_0\rangle$ to $|m_\tau \rangle$  given by  $p^\tau_{m \vert n}=|\la m_\tau|\hat{U}(\tau)|n_0\ra|^2$. The work associated with such a transition is set by the energy difference between the associated eigenvalues, $W=E_m^\tau -E_n^0$. The definition of the work pdf in (\ref{defwork}) encompasses all possible transitions with the corresponding probabilities. As already advanced in the introduction, the need for two projective-energy measurements prevents the introduction of a Hermitian work operator \cite{Talkner2007a}. However, it is possible to describe work in terms of a positive-operator value measure (POVM)  \cite{Mazzola2013a,Roncaglia2014a}.
\\
\\
\emph{Work in chaotic quantum systems.---}
Quantum chaotic systems are described by random matrix Hamiltonians that have been extensively studied following the seminal works by Wigner and Dyson  \cite{MehtaBook,ForresterBook}. 
We here focus on the quantum work statistics associated with the driving of a system between two chaotic Hamiltonians, each independently sampled from  the Gaussian unitary ensemble (GUE), i.e., described by random $N \times N$ Hermitian matrices invariant under  conjugation by unitaries. Here, $N$ is the dimension of the Hilbert space that can be related to the size of a $N_p$ particle system (for qubits, $N=2^{N_p}$).

We shall consider the initial state to be described by the canonical density matrix of a thermal state at inverse temperature $\beta= (k_B T)^{-1}$,  
\beqa
\label{rhothermal}
\hat{\rho}_{\rm th}=\frac{e^{-\beta\hat{H}_0}}{Z(\beta)}\, , 
\eeqa
where  $Z(\beta)=\tr \left(e^{-\beta\hat{H}_0} \right) = \sum_n e^{-\beta E_n^0}$ is the partition function of the initial Hamiltonian $\hat{H}_0$. 
This density matrix, being diagonal in the energy eigenbasis of the initial Hamiltonian,  is left unchanged by the first projective energy measurement. 
The occupation probability of the $n$-initial eigenmode is thus given by the Boltzmann factor, $p_n^0 = \exp(-\beta E_n^0)/Z(\beta)$ where $E_n^0$ is the $n$-th eigenstate of the initial Hamiltonian $\hat{H}_0$, that is sampled from GUE. The  Hamiltonian at the end of the evolution $\hat{H}_\tau$ is also sampled from the GUE. As a result, we are led to consider the average work pdf $\la p_\tau(W)\ra$ that includes a double GUE average over $\hat{H}_0$ and $\hat{H}_\tau$, 
\begin{equation} \label{pWGUE}
\langle p_\tau(W)\rangle =\left \langle \!\! \left \langle\sum_{n,m} p^0_{n}\;  p^\tau_{m \vert n} \delta\left[W\!-\left(E_m^\tau -E_n^0\right)\right] \right \rangle \!\! \right \rangle. 
\end{equation}

For its study, we shall use the average of an arbitrary operator $\hat{O}$ over the probability density function $\rho(\hat{H})$  from which realizations of the chaotic Hamiltonian $\hat{H}$ are sampled in GUE. This average can be written as    \cite{MehtaBook, Lobejko2017a}
\beqa \label{GUE_avg}
\la \hat{O} \ra &\equiv& \int d\hat{H} \rho(\hat{H})\,\hat{O} = \int D E \, \hat{O}(E)\, , 
\eeqa
where, in terms of the eigenvalues $E_n$ of $\hat{H}$, the GUE measure $DE$ has been defined to  include the eigenvalue Vandermonde determinant $\Delta(E) = \prod_{k>j} (E_k - E_j)\,$  and all the Gaussian terms in the pdf. Therefore, the normalization condition gives the  scaling of eigenvalues with the system size and  reads  $\int DE = C_N\, \int \prod_k dE_k  |\Delta(E)|^2 e^{- \sum_j E_j^2} = 1$, the constant $C_N$ being given in Ref. \cite{MehtaBook,ForresterBook}. 

As we have seen, the measurement of work singles out the energy eigenbasis as a preferred basis. In the following, we shall be able to express all the ensemble averages using the density of states $\rho(E)=\sum_n\, \delta(E - E_n)$. When the Hilbert space dimension $N$ is large, the averaged $\la \rho(E)\ra $ is well described by Wigner's semicircle law \cite{MehtaBook}. 
A common approach in random matrix theory to account for a finite Hilbert space dimension $N$ relies on perturbative expansions in $1/N$. However, in the GUE, the exact  $\la \rho(E)\ra $ for any finite $N$ is simply read \cite{MehtaBook}, 
\beqa
\label{edos}
\langle\rho(E)\rangle=\sum_{j=0}^{N-1}\varphi_j(E)^2\, , 
\eeqa
where  the  $\varphi_j(E)={(2^jj!\sqrt{\pi})^{-1/2}}e^{-\frac{E^2}{2}}\mathcal{H}_j(E)$ are given in terms of the   Hermite polynomials $\mathcal{H}_j$.
Knowledge of the density of states will allow us to characterize the work statistics in systems of arbitrary size. To this end, we shall also need the connected two-level correlation function   \cite{MehtaBook}, 
\beqa
\label{eq.TwoLevelCorrelationHermite}
\left\langle\rho_c^{(2)}(E,E') \right\rangle
&=& \la\rho(E)\rho(E')\ra-\la\rho(E)\ra\la\rho(E')\ra  \nonumber\\
&=& -\left( \sum_{j=0}^{N-1} \varphi_j(E) \varphi_j(E')\right)^2\,,
\eeqa
that accounts for the presence of correlations between different energy levels.

Thus, the results we obtain in this manuscript are nonperturbative in $N$ and account exactly for the finiteness of the Hilbert space dimension, 
which can be varied from small to arbitrarily large values. This is of particular importance to describe the quantum work statistics of finite quantum 
chaotic systems as well as to explore the connections between work fluctuations and information scrambling.

\subsection{Characteristic function and work distribution}

The characteristic function of the work pdf is defined  as its Fourier transform 
\beqa
\chi(t,\tau) =\int_{-\infty}^\infty dW p_\tau(W) \,e^{iWt}\, .
\label{defchi}
\eeqa
Using the definition of $p_\tau(W)$, the explicit form of the characteristic function can be written  as
\beqa
\chi(t,\tau)=\sum_{n}p_n^0\la n_0|e^{it\hat{H}_\tau^{\rm eff}}e^{-it\hat{H}_0}|n_0\ra\, ,
\eeqa
which suggests an interpretation of the variable $t$ as a second time of evolution -- different from the physical time $s$ -- with a quench from the initial Hamiltonian to the effective Hamiltonian \cite{Silva2008a,Garcia-Mata2017a,Garcia-Mata2018a,Chenu2018a} 
\beqa \label{eq:Heff}
\hat{H}_\tau^{\rm eff}=\hat{U}^\dagger(\tau)\hat{H}_\tau \hat{U}(\tau)\, ,
\eeqa
defined in terms of the time-evolution operator $\hat{U}(\tau)$ in Eq. (\ref{Utau}).

In what follows, we aim at deriving an exact analytical solution of the characteristic function valid at finite $N$. We shall focus on the average of this quantity for Hamiltonians sampled from the GUE. 
Specifically, for an initial thermal state $\hat{\rho}_{\rm th}$,  the characteristic function of the work distribution associated with a sudden quench protocol between two GUE Hamiltonians  can be written as 
\beq \label{gfguee}
\left\langle \chi(t,\tau)\right\rangle =
\left\langle \!\! \left\langle \frac{1}{Z(\beta)}{\rm Tr}\left( e^{-\sigma_\tau\hat{H}_\tau}\, e^{-\sigma_0\hat{H}_0}\right) \right\rangle \!\!  \right\rangle\, , 
\eeq 
with $\sigma_0 = \beta + it,\, \sigma_\tau = -it$, and where the brackets denote a double average over the initial and final  GUE ensembles.

 Using the spectral resolution of the initial and final Hamiltonians, we find
\beqa \label{gfguea}
 \left\langle \chi(t,\tau)\right\rangle &=&  \frac{1}{\la Z(\beta)\ra} \times\\
& &\left \la \!\! \left \la\sum_{n,m} |\braket{m_\tau}{n_0}|^2 e^{-\sigma_\tau E_{m}^\tau} e^{- \sigma_0 E_{n}^0}\right \ra \!\!\right \ra\, .\nonumber
\eeqa
Note that in going from (\ref{gfguee}) to (\ref{gfguea}), the average of the fraction has been approximated as the ratio of 
 the numerator and denominator averages. This kind of approximation is known as annealing in the context of random matrix theory. The resulting  `annealed' expression of the characteristic function is required for  an analytical treatment,  and we verify numerically that it is a good approximation.  
Introducing the density of states  $\rho(E)=\sum_n\, \delta(E - E_n)$
and using the average of the transition probability over the  GUE $\left\langle p^\tau_{m\vert n}\right\rangle  = 1/N$ \cite{Lobejko2017a}, we  find
\begin{eqnarray}
 \left\langle \chi(t,\tau)\right\rangle  &=& \frac{1}{N \, \la Z(\beta)\ra }\iint dE_0\, dE_\tau\, \la \rho(E_0) \ra\, \la \rho(E_\tau)\ra\, \nonumber\\ 
& & \times e^{-\sigma_\tau E_\tau}\, e^{-\sigma_0 E_0}\, . 
\end{eqnarray}
A compact form can be derived by defining the Laplace transform of the density of states averaged over a GUE, 
\beqa
\label{LaplaceDOS}
\mathcal{I}(\sigma)= \int dE\, \la \rho(E) \ra\, e^{-\sigma E}\, . 
\eeqa
It is worth pointing out that the related real Fourier transform of the density of states is used as a tool to measure statistical level properties of systems with a complex energy spectrum  \cite{Leviandier1986a,Wilkie1991a,Alhassid1993a,Ma1995a}.

The explicit evaluation of (\ref{LaplaceDOS}) for finite  $N$ requires the expression for the GUE density of states (\ref{edos}), 
and leads to 
\beqa \label{eq:I_def}
\mathcal{I}(\sigma)=e^{\frac{\sigma^2}{4}}L_{N-1}^1\left(-\tfrac{\sigma^2}{2}\right)  = \la Z(\sigma) \ra\,,
\eeqa 
where $L_n^{\alpha}(x) = \sum_{i=0}^{n} \binom{n+\alpha}{n-i} (-x)^i / i !$ denotes the generalized Laguerre polynomial.  Physically, $\mathcal{I}(\sigma)$ corresponds to the average of the  analytically continued partition function  \cite{delCampo2017a}. 
The GUE averaged partition function is readily obtained as,   
 \beqa \label{eq:ZI}
 \la Z(\beta)\ra =\mathcal{I}(\beta)\xrightarrow[\beta \rightarrow 0]{} L^1_{N-1}(0) = N\, ,
 \eeqa
 where we verify that in the infinite-temperature limit the partition function measures the Hilbert space dimension $N$.
Thus, we obtain  the following compact and explicit expression for the characteristic function 
 \begin{flalign} \label{eq:chi}
 &\left\langle \chi(t,\tau)\right\rangle =
 \frac{1}{N}\frac{ \la Z(\sigma_0) \ra \la Z(\sigma_\tau)\ra }{ \la Z(\beta) \ra}   &\\
 &= \frac{1}{N} e^{\frac{1}{2} (\beta^2 - t^2) } e^{i \frac{\beta t}{2}} \frac{L^1_{N-1}(-\frac{(\beta + it )^2}{2}) L^1_{N-1}(\frac{t^2}{2})}{L^1_{N-1}(-\frac{\beta^2}{2})}\, .\nonumber 
 \end{flalign}
 Note that the characteristics function in the limit of large $N$ has also been studied in \cite{Arrais2018a}.
 
\begin{figure}[t]
  \includegraphics[width=1\columnwidth]{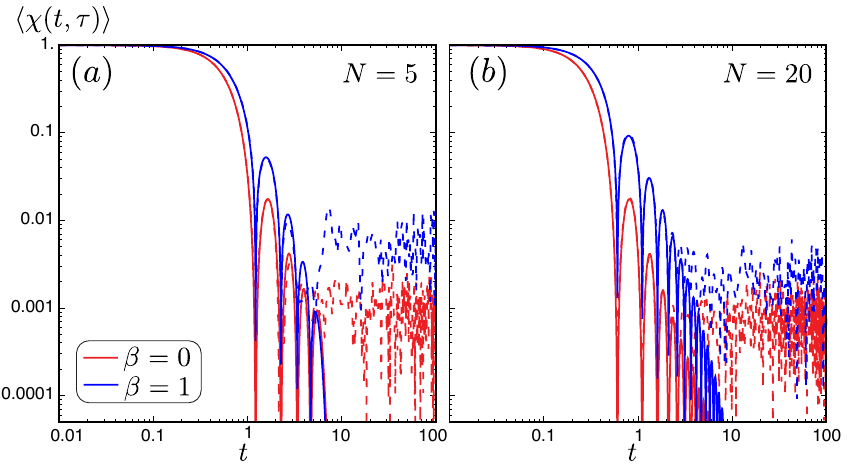}
 \caption{{\bf Characteristic function of the work pdf for a sudden quench between two GUE Hamiltonians.} $\la \chi(t,\tau)\ra$ from the analytical solution  (solid line, Eq. \ref{eq:chi}), and as obtained numerically averaging over 50000 realizations of the GUE at infinite temperature (red) and $\beta=1$ (blue). The numerical solutions converge very slowly and tend to 0 at large time. The dimension of the Hilbert space is (a) $N=5$ and (b) $N=20$.
  \label{fig:chi}}
 \end{figure}

In the infinite temperature limit, it simplifies to $ \left\langle \chi(t,\tau)\right\rangle=e^{-t^2 /2} [L^1_{N-1}(t^2/2)]^2 / N^2$ and can be written in the form of a Gaussian multiplied by a polynomial of even powers up to $2(N-1)$. 
A  log-log representation exhibits a number of dips that corresponds to the positive roots. The number of dips is thus $(N-1)$, before a final monotonic decay. Figure~\ref{fig:chi} presents the analytical results for different temperatures and Hilbert space dimensions. It illustrates the importance of the closed-form expression (\ref{eq:chi}) over the numerical simulations that exhibit poor convergence properties. 
 
\begin{figure}[t]
 \includegraphics[width=1\columnwidth]{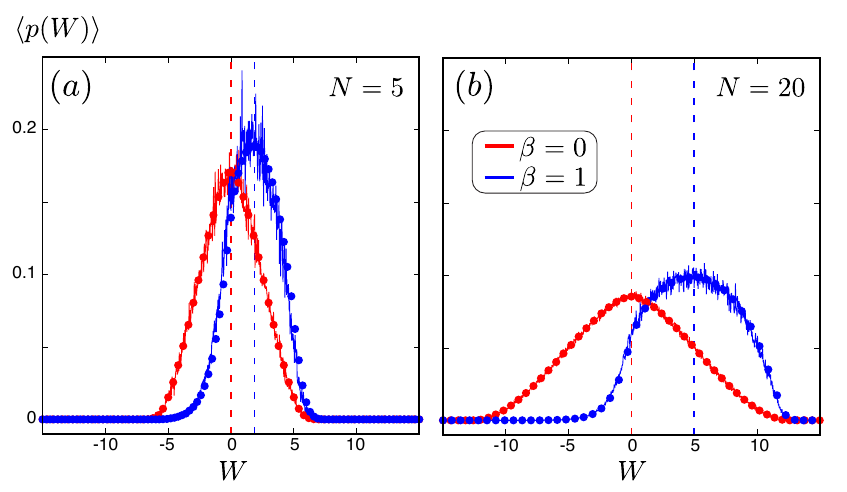}
 \caption{{\bf Work statistics in a  chaotic quantum system driven by a sudden quench.} Work pdf $\la p(W)\ra$ obtained from the analytical expression Eq. (\ref{eq:pWGUE}) (circles), and  the numerical solution averaging over 50000 realizations of the GUE (solid lines), at infinite temperature (red) and $\beta=1$ (blue). The vertical dashed lines mark the corresponding average work given in Eq. (\ref{meanw}). We verify the symmetric form at infinite temperature, yielding an average work $\la W\ra =0$. The dimension of the Hilbert space is (a) $N=5$ and (b) $N=20$. \label{fig:pWAvg}}
 \end{figure}

 From the characteristic function in Eq. (\ref{eq:chi}),  the average work probability distribution  can be obtained by inverting the Fourier transform (\ref{defchi}). The work pdf is found to be given by  the auto-convolution of the average density of states $\la \rho(E) \ra$ with an exponential weight, 
\beq \label{eq:pWGUE}
\la p(W)\ra =\frac{1}{N \mathcal{I}(\beta)}  \int \limits_{-\infty}^{\infty}  dE \la \rho(E) \ra \la \rho(E{+}W) \ra e^{{-}\beta E}\, .
\eeq
We present the average work pdf in Figure~\ref{fig:pWAvg}, and verify that results obtained from the numerical integration of the above equation coincide with those obtained from direct sampling over many realizations of the GUE.  In the infinite temperature limit, the even parity of the density of states $\la\rho(E)\ra=\la\rho(-E)\ra$ carries over the work pdf, $\la p(W)\ra=\la p(-W)\ra$. As a result, any odd moment of the distribution $\la W^n\ra=\int dW  p(W)W^n$,  including the  mean work ($n=1$),  vanishes identically in this limit.  Away from it, the work pdf is described by a bell-shaped function whose mean shifts from zero to higher positive values as the temperature of the initial canonical state is decreased.
 
\subsection{Mean work in a random quench}
The first moment of the work pdf, the mean work, is of both practical and fundamental interest. It controls the performance of quantum thermodynamic devices such as heat engines and refrigerators. In addition, it can be used to assess irreversibility and inner friction of quantum thermodynamic processes \cite{Deffner2010a,Plastina2014a}.

\begin{figure}
\includegraphics[width=1\columnwidth]{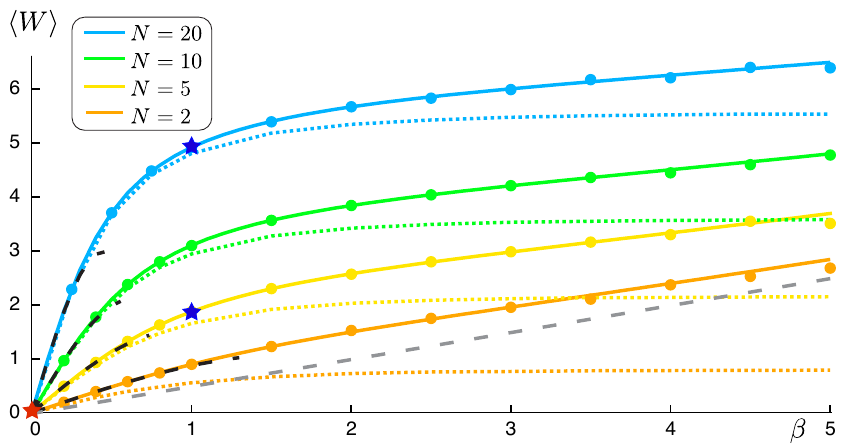}
\caption{ 
{\bf Temperature dependence of mean work.} For  chaotic systems of various Hilbert space dimension $N$, the mean work  $\la W \ra$ associated with a sudden quench is shown as function of the inverse temperature $\beta$. Both the annealed expression  in Eq. (\ref{meanw})  (solid lines) and numerical   results (dots) are plotted. Comparison with the numerical results without annealing (dotted lines) shows the importance of the latter, as discussed in the text. The dashed grey line shows the asymptote $\beta/2$ at low temperatures; the dashed black lines correspond to the asymptotes at high temperatures, as given in Eq. (\ref{eq:W_smallbeta}). The  star symbols indicate the conditions for which the detailed  characteristic function and work pdf are given in  Figs. \ref{fig:chi} and \ref{fig:pWAvg}. \label{fig:W}}
\end{figure}

The mean of the work pdf is directly given by the difference between the initial and final mean energies, 
\beqa
\la W\ra&=&\la \hat{H}_\tau \ra-\la \hat{H}_0\ra\\
&=&\left\la \!\! \left\la\tr\left[(\hat{H}_\tau-\hat{H}_0)\frac{e^{-\beta \hat{H}_0}}{\tr\left(e^{-\beta \hat{H}_0}\right)}\right]\right\ra \!\!\right \ra\, , \nonumber
\eeqa
which involves a double GUE average over $\hat{H}_0$ and $\hat{H}_\tau$.
The mean final energy vanishes identically, i.e., 
\beqa
\la \hat{H}_\tau \ra &=&\left\la \!\!\left\la \tr\left[\hat{H}_\tau \frac{e^{-\beta \hat{H}_0}}{\tr\left(e^{-\beta \hat{H}_0}\right)}\right]\right\ra \!\!\right\ra = 0\, , 
\eeqa
because of the spectral symmetry of the final ensemble. 
The  mean work is thus set exclusively by the initial mean energy and reads
\beqa
\label{meanw}
\la W\ra=-\, \la \hat{H}_0\ra 
 &=& \frac{1}{\la Z(\beta)\ra}\frac{d \la Z(\beta)\ra} {d\beta} \nonumber \\
&=&\frac{\beta}{2}+\beta\, \frac{L_{N-2}^2(-\beta^2/2)}{L_{N-1}^1(-\beta^2/2)}\, .
\eeqa
For high temperature ($\beta\ll1$), the exact asymptotic behavior is thus
\beqa  \label{eq:W_smallbeta}
\la W\ra \xrightarrow[\beta \rightarrow 0]{}\frac{N}{2}\beta-\frac{N^2-1}{24}\beta^3+\mathcal{O}(\beta^5)\, , 
\eeqa
verifying the fact  that the mean work  vanishes exactly in the infinite  temperature limit and increases as the temperature is decreased, as seen in Figure~\ref{fig:pWAvg}. 
The mean work is lower-bounded by the low-temperature limit, 
\beqa \label{W_betaInf}
 \la W\ra \xrightarrow[\beta \rightarrow \infty]{} \frac{\beta}{2}\, ,
\eeqa
%\eeqa
as  in this regime the second term in the r.h.s. of Eq. (\ref{meanw}) vanishes.
The behavior of the mean work for chaotic systems as a function of the inverse temperature is presented in Figure~\ref{fig:W} for different dimensions of the Hilbert space  and illustrates the accuracy of the  high temperature asymptotes.

  Comparing the analytical results with the numerical simulations, we find discrepancies between the `annealed' and exact expression at low temperatures. 
 This can be understood by the shape of the distribution of the GUE Hamiltonians that, as the temperature lowers, has an increasing non-zero  tail that eventually requires more sampling---in contrast with the large $N$ limit where the semi-circle distribution ensures a finite sampling domain. Physically, the work, set by  (the negative of) the ensemble initial energy, should be bounded at low temperature. This is clearly not the case in the analytical asymptote in Eq. (\ref{W_betaInf}), which shows the limitations of the annealed approximation made in Eq. (\ref{gfguea}) in the low temperature regime, needed for the pursued analytical treatment.

\subsection{Variance of the work pdf in a random quench}
The stochastic nature of work in the presence of quantum thermal fluctuations becomes manifest when considering fluctuations away from the mean $\la W\ra$. 
We focus on  the variance that naturally characterizes the width of the work pdf. As we shall see, it also governs the decay of the Loschmidt echo at early times of evolution \cite{Chenu2018a} and constitutes the leading contribution to the irreversible entropy in linear response theory \cite{Goold2018a}.

\begin{figure}

\includegraphics[width=1\columnwidth]{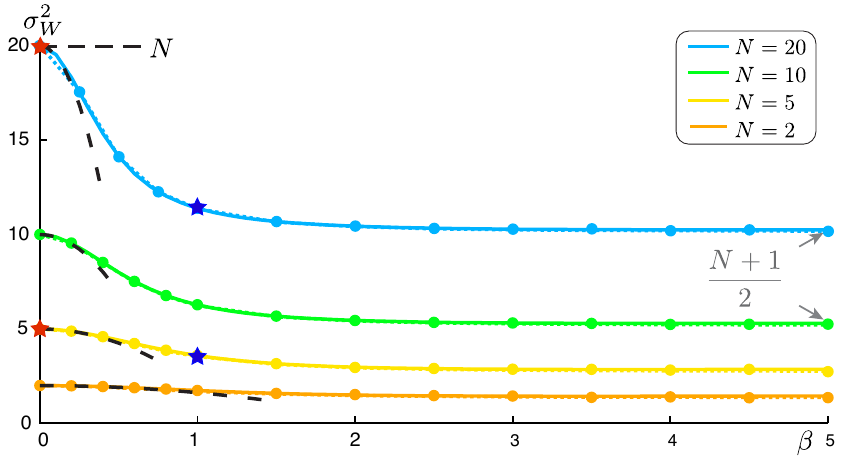}
\caption{ 
{\bf Temperature dependence of work fluctuations.}  Variance of the work $\sigma_W^2$ given in Eq. (\ref{DW2}) for different system sizes as function of the inverse temperature $\beta$. The fluctuations monotonically decay from the infinite-temperature value $N$ following the dashed-line asymptote corresponding to Eq. (\ref{sigma_asymptote0}),  saturating at $(N+1)/2$ in the low-temperature regime.  The analytical results (solid lines) are in excellent agreement with the numerical simulations for both the annealed  (circles) and the exact   (dotted lines) expressions.  \label{fig:sigmaW}}
\end{figure}

An explicit computation shows that the second moment of the work pdf simplifies to
\begin{eqnarray} \label{eq:W2_def}
\la W^2 \ra &=&\left\la \int_{-\infty}^\infty dW p(W) \, W^2 \right \ra  \\
&=&  \left \la \!\!\left \la \tr\! \left( (\hat{H}_\tau^2 + \hat{H}_0^2 ) \frac{e^{-\beta \hat{H}_0}}{\tr(e^{-\beta \hat{H}_0})}\right)\! \right\ra   \!\!\right\ra\, ,\nonumber
\end{eqnarray}
as a result of  the  aforementioned symmetry in the final ensemble, that yields 
 $\la\! \la \hat{H}_\tau \hat{H}_0\ra\! \ra=0$. 
The GUE-averaged expectation value of the second moment for the initial Hamiltonian can be derived from the partition function in Eq. (\ref{eq:ZI}),  
\begin{equation}
\label{hi2}
\left \la \tr \left( \hat{H}_0^2 e^{-\beta \hat{H}_0} \right)  \right \ra= \frac{d^2  \la Z(\beta) \ra}{d\beta^2}\, . 
\end{equation}
By contrast, the second moment of the final Hamiltonian requires averaging over both the initial and final state ensembles, specifically 
 \begin{flalign}
\label{hf2}
& \left \la\!\! \left \la \tr \left( \hat{H}_\tau^2 \, e^{-\beta \hat{H}_0} \right)   \right \ra\!\! \right \ra & \nonumber\\ 
&= \left \la \!\!\left \la \sum_{m,n} |\braket{m_\tau}{n_0} |^2 \! \left(E^\tau_m \right)^2 \! e^{-\beta E^0_n}    \right \ra \!\! \right \ra \nonumber \\
&=\la p_{m|n}^\tau \ra  \left \la \sum_m \left( E^\tau_m\right)^2  \right \ra  \left \la \sum_n e^{-\beta E^0_n}   \right \ra \nonumber \\
&=  \frac{1}{N} \left. \frac{d^2 \mathcal{I}(\beta)}{d\beta^2} \right\vert_{\beta=0} \la Z(\beta) \ra \nonumber \\
&= \frac{N}{2} \la Z(\beta) \ra\,   .
\end{flalign}
Equations (\ref{eq:W2_def})-(\ref{hf2}), along with the definitions (\ref{eq:I_def})  and (\ref{eq:ZI}),  give the   exact, non-perturbative expression for the work fluctuations  as 
\begin{eqnarray}
\label{w2}
\la W^2 \ra &=&   \frac{N}{2}  +   \frac{1}{\la Z(\beta) \ra}\frac{d^2 \la Z(\beta) \ra }{d\beta^2} \nonumber \\
&=&\frac{N{+}1}{2}  + \frac{\beta^2}{4}   + (1{+}\beta^2) \frac{L_{N-2}^2 \left(-\frac{\beta^2}{2} \right)}{L_{N-1}^1\left(-\frac{\beta^2}{2} \right)} \nonumber \\
&&+ \beta^2 \frac{L_{N-3}^3\left(-\frac{\beta^2}{2} \right)}{L_{N-1}^1\left(-\frac{\beta^2}{2} \right)}\, . 
\end{eqnarray}
Knowledge of  the first two moments of the distribution, (\ref{meanw}) and (\ref{w2}), readily yields the work variance   
\begin{eqnarray} \label{DW2}
 \sigma_W^2
&=& \la  W^2\ra -\la  W\ra^2 \nonumber \\
&=& \frac{N+1}{2}+ \frac{L_{N-2}^2\left(-\frac{\beta^2}{2}\right)}{L_{N-1}^1\left(-\frac{\beta^2}{2}\right)}     \\
&&+ \beta^2\left[ \frac{L_{N-3}^3\left(-\frac{\beta^2}{2}\right)}{L_{N-1}^1\left(-\frac{\beta^2}{2}\right)} - \left( \frac{L_{N-2}^2\left(-\frac{\beta^2}{2}\right)}{L_{N-1}^1\left(-\frac{\beta^2}{2}\right)} \right)^2 \right]\, .  \nonumber
\end{eqnarray}
At infinite temperature, the work variance saturates at the Hilbert space dimension $N$ following the asymptotic behavior
\beqa \label{sigma_asymptote0}
\sigma_W^2  \xrightarrow[\beta \rightarrow 0]{} N-\frac{N^2-1}{8}\beta^2+\mathcal{O}(\beta^4)\, . 
\eeqa
In the low temperature regime, the work variance is still finite and saturates at the value  of $(N+1)/2$ beyond $\beta \gtrapprox 1.5$. 
Both limits are presented in Figure~\ref{fig:sigmaW}, where the monotonic decay of $\sigma_W^2$ is shown as function of $\beta$ for different  values of the Hilbert space dimension. 
 Here, the analytical and numerical results for the annealed approximation match the exact simulation, as the error observed on the average energy (see Figure~\ref{fig:W}) is exactly compensated by that on the energy second moment also used to obtain the variance.

\section{Loschmidt echo in Chaotic Systems and information scrambling}\label{SecLEIS}

The study of the work statistics can be recast into a dynamical problem, since the characteristic function of the work pdf can be related to a  Loschmidt echo. 
A Loschmidt echo quantifies the degree to which a given dynamics can be reversed \cite{Gorin2006a,Goussev2012a}.  
The connection between work statistics and Loschmidt echoes was first established for a system initialized  in an energy eigenstate and driven by a sudden quench  \cite{Silva2008a}. Following works extended this relation to  arbitrary evolutions by interpreting  the characteristic function  as a Loschmidt echo amplitude where the backward evolution is generated by an effective Hamiltonian \cite{Garcia-Mata2017a,Garcia-Mata2018a}. More recently, we have shown that such a connection holds  as well for an arbitrary initial state, whether it is pure or not \cite{Chenu2018a}. In turn,  the ability to recast work statistics in terms of a dynamical problem (a Loschmidt echo) for any arbitrary initial states results is an important connection between work statistics and information scrambling.

To appreciate this, we first consider a general initial state $\hat{\rho}$, which may include coherences in the initial energy eigenbasis. Following the first energy measurement associated with the projectors $\hat{P}^0_n=|n_0\rangle\langle n_0|$,  the  post-measurement state  becomes $\sum_n \hat{P}^0_n \hat{\rho}  \hat{P}^0_n=\sum_{n} p^0_n\, |n_0\rangle\langle n_0|$. This mixed state can be purified by embedding it in an extended Hilbert space, e.g.,  $\mathcal{H}_L\otimes\mathcal{H}_R$, with $\mathcal{H}_L = \mathcal{H}_R =\mathcal{H}$. The purified state  is then given by an entangled state between two copies of the system 
\beqa
\label{pstate}
|\Psi_0\ra=\sum_n\sqrt{p^0_n}\,
\,|n_0\ra_L\otimes|n_0\ra_R\, . 
\eeqa

 The characteristic function associated with the evolution $\hat{U}(\tau)$ in  Eq. (\ref{Utau}) can then be  expressed as  the survival amplitude $\la\Psi_0|\Psi_t\ra$ of the purified state (\ref{pstate}) when one of the copies of the system evolves under the sudden quench 
\begin{equation} \label{genquench}
\hat{H}(t)=\hat{H}_0\, \Theta(-t)-\hat{H}_\tau^{\rm eff}\, \Theta(t)\,,
\end{equation}
this is, a sudden quench between the initial Hamiltonian and the effective Hamiltonian $-\hat{H}_\tau^{\rm eff}$ defined in Eq. (\ref{eq:Heff}),  $\Theta(t)$ denoting the Heaviside function. Specifically, the equivalence becomes explicit by composition of the unitary evolution operators,  and reads  \cite{Chenu2018a}
\beqa
\label{chigen}
\chi(t,\tau)&\equiv&\la\Psi_0|\Psi_t\ra \nonumber  \\
&=&\la\Psi_0|  e^{+it\hat{H}_\tau^{\rm eff}}e^{-it\hat{H}_0}  \otimes \mathbb{1}_R |\Psi_0\ra\, ,
\eeqa
where we emphasize that the quench evolution acts exclusively on one of the copies, e.g., the left one. Thus, the characteristic function in Eq. (\ref{chigen}) also encodes  the dynamics of the Loschmidt echo amplitude of the pure state $|\Psi_0\rangle$ in which the left copy evolves first forward in time under the Hamiltonian $\hat{H}_0$ and then backward under the effective Hamiltonian  $\hat{H}_\tau^{\rm eff}$, leaving  the right copy unchanged. 

We stress here that the connection $ \chi(t,\tau)\equiv \la\Psi_0|\Psi_t\ra$ is valid  for  any driving protocol and for any purified state $|\Psi_0\rangle$, including those arising from mixed states at finite temperature \cite{Chenu2018a}.  One may thus consider the initial state to be the canonical thermal state, as in  Eq. (\ref{rhothermal}).
 In this particular case, the corresponding purified state  is the  so-called thermofield double state \cite{Semenoff1983a}, 
\beqa
\label{TDS}
\ket{{\Phi}_0 }=\frac{1}{\sqrt{Z(\beta)}}\! \sum_ne^{-\frac{\beta}{2} \hat{H}_0\,\otimes\, \mathbb{1}_R }\,|n_0\ra_L\otimes|n_0\ra_R\, .\nonumber\\
\eeqa
Its time-evolution is then described by
\beq
\label{TFD_echo}
\ket{ \Phi_t } = e^{-i \int_0^t dt'\hat{H}(t')\, \otimes\, \mathbb{1}_R  }\, |\Phi_0 \rangle\, ,
\eeq
with $\hat{H}(t')$ given in Eq. (\ref{genquench}).  Its survival probability amplitude is equivalent to the characteristic  function of the corresponding work statistics, namely 
\beqa
\label{chiTDS}
\chi(t,\tau) & = & \braket{\Phi_0 \,}{ \, \Phi_t} \\ \nonumber
&=&\frac{1}{Z(\beta)}\, {\rm Tr}\left(  e^{+it\hat{H}_\tau^{\rm eff}}e^{-(\beta +it)\hat{H}_0}\right)\, . 
\eeqa
In this interpretation, the thermofield double state (\ref{TDS}) is not invariant under time-evolution and the equivalence established above gives the  Loschmidt echo  $\mathcal{L}(t)$ as the fidelity between the initial thermofield double state $|\Phi_0\ra$ and its time-evolution, also known as the survival probability:
\beqa
\label{letds}
\mathcal{L}(t)=| \braket {\Phi_0 \,} {\, \Phi_t} |^2\, =\left|\int_{-\infty}^\infty dW p_\tau(W) \,e^{iWt}\right|^2\, .\nonumber\\
\eeqa
Said differently, the Loschmidt echo of the thermofield double state can be obtained from the Fourier transform of the work pdf.
This observation leads to a direct connection between quantum thermodynamics -- the analysis of work statistics -- and information scrambling, as characterized by $\mathcal{L}(t)$.

Information scrambling was first discussed in black hole physics \cite{Hayden2007a,Sekino2008a,Barbon2003a,Barbon2004a}. In particular, within the AdS/CFT correspondence, an eternal black hole, or so-called Einstein-Rosen bridge, is dual to the  thermofield double state, and described by a state with a highly localized entanglement distribution between the degrees of freedom of the two non-interacting copies of a CFT \cite{Maldacena2003a}. The notion of information scrambling refers to the redistribution of any quantum information, initially localized in a subsystem, over the entire system under unitary time evolution. Thus, in the AdS/CFT framework, it is natural to study the spreading of information in black holes via the survival probability of a thermofield double state, that is, via the fidelity between the initial state and its unitary time evolution, as in Eq. (\ref{chiTDS}) \cite{Papadodimas2015a,Miyaji2016a, Cotler2017b, Dyer2017a, delCampo2017a}. The time-evolved state $|\Phi_t \rangle$ in Eq. (\ref{TFD_echo}) can be understood, in the  gravitational picture, as an   Einstein-Rosen bridge which interior has grown  \cite{Maldacena2013a,Susskind2016a}. This notion of information scrambling has proved useful in a more general setting, in which initially localized information in a subsystem spreads and gets scrambled in the degrees of freedom of a many-body system. In the following, we characterize this dynamics in  chaotic quantum systems represented by the Gaussian unitary ensemble.

\subsection{Loschmidt Echo in GUE} 
The Loschmidt echo, defined in Eq. (\ref{letds}), averaged over the GUE follows from the characteristic function in Eq. (\ref{gfguee}) as
\begin{eqnarray} \label{L_def}
\la \mathcal{L} (t)\ra&=& \Big\la\!\! \Big\la\frac{1}{\tr\left(e^{-\beta \hat{H}_0}\right)^2}
\tr\left(e^{-\sigma_\tau \hat{H}_\tau^{\rm eff}}e^{-\sigma_0\hat{H}_0}\right)\quad \nonumber\\
 & & \times
 \tr\left(e^{-\sigma_\tau^*\hat{H}_\tau^{\rm eff}}e^{-\sigma_0^*\hat{H}_0}\right)
 \Big\ra \!\! \Big\ra\, . 
\end{eqnarray}
Using the property of the GUE invariance under conjugation by a unitary, the double average reduces to two independent GUE averages, one over $\hat{H}_0$ and one over $\hat{H}_\tau$. The `annealed' version of the Loschmidt echo is thus  
\begin{eqnarray} \label{L_annealed}
\la \mathcal{L} (t)\ra
=\frac{1}{\la Z(\beta)^2\ra }\left\la \!\! \left\la
\left|  \tr\left(e^{-\sigma_\tau \hat{H}_\tau}e^{-\sigma_0\hat{H}_0}\right) \right|^2\right\ra \!\! \right\ra\, .\nonumber\\
\end{eqnarray}
Its  explicit form can be conveniently written in terms of the  partition function and the GUE spectral form factor  defined as 
\beqa\label{eq:g}
g(\beta,t) \equiv \left \la Z(\beta+it) Z(\beta - it) \right \ra\, .
\eeqa
 As shown in Appendix \ref{AppA}, it is found that
\beqa \label{L_g}
& & \la \mathcal{L} (t)\ra = \frac{1}{\la Z(\beta)^2\ra }\, \frac{1}{N^2{-}1} \Big( g(0,t)g(\beta,t)  \\
& &  + N \langle Z(2\beta )\rangle- \frac{1}{N}  \langle Z(2\beta )\rangle g(0,t) - \frac{1}{N} g(\beta,t) N \Big)\, .\nonumber
\eeqa
The dependence of the Loschmidt echo on the spectral form factor illustrates the importance of correlations between energy eigenvalues. 
In the infinite-temperature regime ($\beta =0$), this result simplifies to
\beqa
\label{LE_Frame_Pot}
\la \mathcal{L} (t) \ra=\frac{1}{N^2}\, \frac{1}{N^2{-}1} \Big( g(0,t)^2 {+}N^2 {-} 2 g(0,t)\Big)\, .\nonumber\\
\eeqa
The Loschmidt echo can thus be fully characterized from the GUE spectral form factor (\ref{eq:g}), for which we give an explicit formulation below.

\subsection{Loschmidt echo and the spectral form factor}

\begin{figure*}[t]
\includegraphics[width=2\columnwidth]{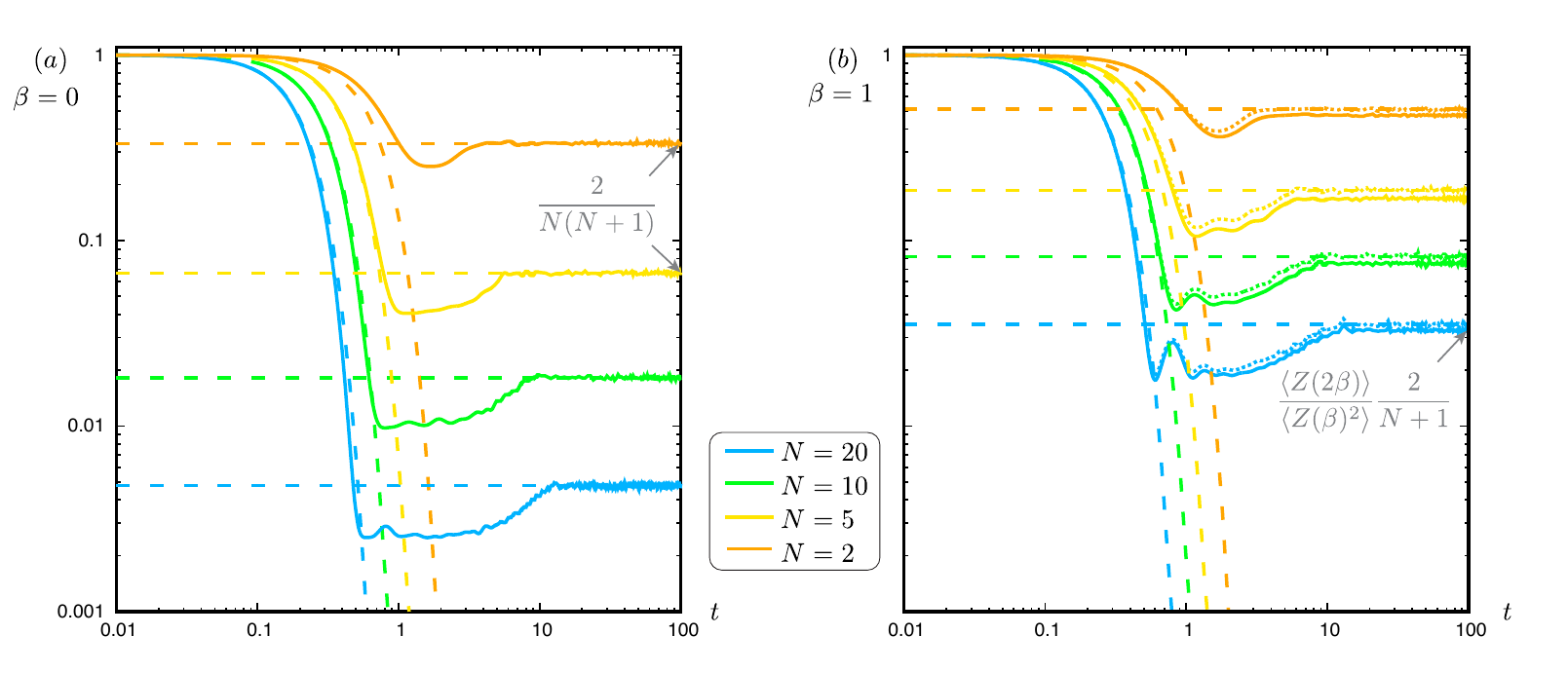}
\caption{{\bf Decay of the Loschmidt echo in GUE.} The time-evolution of the Loschmidt echo is shown  for (a)   $\beta=0$ and  (b) $\beta=1$  and various system sizes $N$.  Three different stages are clearly distinguished in the decay dynamics.  The short-time behavior  is governed by a Gaussian decay dictated by the work variance according to (\ref{Lshort}), see the dashed lines with $ e^{-t^2 N}$ for $\beta=0$. The ensuing dynamics is characterized by a broad dip, and a slow increase towards the plateau at which the Loschmidt echo saturates at long times. The long-time asymptotic value predicted by  Eq. (\ref{L_tinfty}) is marked by the horizontal dashed lines. The figure also shows the difference between the `annealed'  (solid lines) and exact  (dotted lines)  Loschmidt echo, which are indistinguishable in the infinite temperature limit.  
\label{L_asym}}
\end{figure*}

We have shown that the  averaged value of $\mathcal{L}(t)$ can be expressed   in terms of the spectral form factor, as seen in Eq. (\ref{L_g}).  The latter admits an exact analytical expression. Indeed, the form factor  can be decomposed into three contributions as \cite{delCampo2017a} 
\begin{eqnarray} \label{eq:g_def}
g(\beta, t) &=& \la Z(\beta + it) Z(\beta - it)\ra  \\
&=&\langle Z(2\beta) \rangle +|\langle Z(\beta + it)\rangle|^2+g_c(\beta, t)\, , \nonumber
\eeqa
where $g_c(\beta, t)$ is the double complex Fourier transform of the connected contribution, 
\beqa\label{gConnected}
g_c(\beta, t) &=& \int dE dE' \left\langle \rho_c^{(2)} (E,E')\right\rangle\nonumber\\
& & \times e^{- (\beta + it) E - (\beta  - i t)  E'}\, .
\eeqa
The analytically-continued  partition function averaged over the GUE is known from \cite{delCampo2017a} and  given in Eq. (\ref{eq:I_def}).
Similarly, the connected two-level correlation function $\left\langle \rho_c^{(2)} (E,E')\right\rangle=\la\rho(E)\rho(E')\ra-\la\rho(E)\ra\la\rho(E')\ra$  is given by Eq. (\ref{eq.TwoLevelCorrelationHermite}).
Its complex Fourier transform, Eq. (\ref{gConnected}), 
is evaluated in Appendix \ref{AppB} following \cite{delCampo2017a} and reads 
\beqa\label{eq.connectedFormFactorGUE}
g_c(\beta, t)& =&  -  e^{\frac{\sigma_0^2+ \sigma_0^{* 2}}{4} }  \sum_{n,m=0}^{N-1}  c_{m,n} \left( \frac{|\sigma_0|^2}{2} \right)^{|n-m|} \nonumber\\
& &\times \left| L_{\min(m,n)}^{|n-m|}\left(-\frac{\sigma_0^2}{2}\right)  \right|^2\,  , 
\eeqa
where $c_{m,n} = \min(m,n)! / \max(m,n)! $.
 We verify that at infinite temperature and initial time ($\sigma_0=0$), the only non-zero terms in the sum bring a Kronecker delta $\delta_{n,m}$, while the Laguerre polynomials reduce to binomial coefficients, yielding $g_c(0, 0) = -N$ and $\la Z(0)^2 \ra = g(0,0) = N^2$, as expected.

\subsection{Loschmidt echo dynamics}

With the exact analytical expressions at hand, we proceed to characterize the dynamics of the Loschmidt echo. 
The Loschmidt echo is shown in Figure~\ref{L_asym} in two different panels corresponding to infinite temperature ($\beta=0$) and a finite value ($\beta=1$). 
In both cases, the  decay dynamics is shown for different values of the Hilbert space dimension, and one can distinguish a sequence of three  stages. The Loschmidt echo first decays monotonically until reaching a minimum value. The second stage is characterized by a broad dip in which the Loschmidt echo stabilizes around $1/N^2$ and gradually starts to grow. 
Subsequently,  a saturation of the Loschmidt echo occurs around an approximately constant value.

The short time behavior of the Loschmidt echo was first given in our recent work \cite{Chenu2018a} and is governed by work fluctuations -- the variance of the work pdf -- according to
\begin{equation}
\label{Lshort}
\la \mathcal{L}(t) \ra = \big \la e^{-t^2 \sigma_W^2  +\mathcal{O}(t^4) } \big \ra \geq e^{-t^2 \la \sigma_W^2\ra + \mathcal{O}(t^4)}\, ,
\end{equation}
where we have used Jensen's inequality.
For $\beta~=~0$, the work variance given in Eq. (\ref{DW2}) simplifies to $N$, leading  to a Gaussian decay that is simply given by $ e^{-t^2 N}$, as verified in Figure~\ref{L_asym}. 

As for the long-time asymptotics,  it is best analyzed in terms of the spectral form factor. The latter can be written, using Eq. (\ref{eq:g_def}),   as 
\begin{eqnarray}
g(\beta, t) 
&=& \la Z(2\beta)\ra  \\
& & +\left  \la \sum_n \sum_{n' \neq n} e^{-\beta (E_n + E_{n'}) } e^{- it (E_n - E_{n'})} \right \ra\,  . \nonumber
\end{eqnarray}
At long times, the complex exponential  exhibits rapid oscillations and only the first term survives under a short time average that enters in the feature of a  plateau. 
Taking the limit $g(\beta,t) \to  \la Z(2\beta)\ra $, Eq. (\ref{L_g}) gives the asymptote of the Loschmidt echo at long times as
\begin{equation} \label{L_tinfty}
\la \mathcal{L}(t) \ra \xrightarrow[t\to \infty]{} \frac{\la Z(2\beta) \ra}{\la Z(\beta)^2 \ra}  \frac{2}{(N+1)} 
\end{equation}
for the `annealed' expression, Eq. (\ref{L_annealed}). We numerically validate the approximation splitting the numerator and denominator averages and compare with the exact expression, defined in Eq. (\ref{L_def}), which is found  slightly below this asymptote with a limit given by  $\left \la \frac{ Z(2\beta) }{ Z(\beta)^2 } \right \ra  \frac{2}{(N+1)}$. 
At infinite temperature, $\beta=0$, the exact and annealed asymptotes are identical and equal to $2/(N (N+1))$.

\subsection{Loschmidt Echo, frame potentials and scrambling}\label{AppC}

As discussed in Section \ref{SecLEIS}, quantum work statistics can be used to assess information scrambling, as the generating function of the work pdf is equivalent to the Loschmidt echo amplitude of a thermofield double state.
Scrambling amounts to the effective randomization of the state of a quantum system subject to unitary time evolution. Depending on the dynamics, scrambling produces different degrees of randomness. There is a wide variety of approaches to pursue its study, including the use of survival probabilities, analytically-continued partition functions, and out-of-time-order correlators, among other examples. 

In black hole physics, the common reference is the Haar/Page information scrambling, which consists in drawing Haar unitaries  from the Haar ensemble. This yields a complete randomization of a quantum state, making it  impossible to recover information about the initial state by carrying out only local measurements on the system. In this context, it is  relevant to analyze the extent to which a concrete ensemble of unitaries can match Haar scrambling. 

Frame potentials have been argued to act in this context as natural probes of the randomness created by a concrete time evolution, by quantifying an ensemble ability to reproduce moments of the Haar random unitary ensemble \cite{Cotler2017a}. 
The $k$-th frame potential with respect to an ensemble $\mathcal{E}$ is defined as 
\beqa
 \mathcal{F} ^{(k)}_{\mathcal{E}} =  \int_{\hat{A},\hat{B}\, \in \mathcal{E}} D\hat{A}\, D\hat{B} \left\vert\, \tr\, \hat{A}^{\dagger}\, \hat{B}\, \right\vert^{2k}\, .
\eeqa
For any ensemble, $ \mathcal{F} ^{(k)}_{\mathcal{E}} \geq \mathcal{F} ^{(k)}_{\rm Haar}$, 
where the Haar ensemble frame potential is $\mathcal{F} ^{(k)}_{\rm Haar} =k!$.

In our case, the relevant ensemble consists of the unitary time evolutions at a fixed time $t$ with Hamiltonians drawn from the GUE and reads  
%\beqa
$
\mathcal{E}_t^{\rm GUE}=\big\lbrace e^{-i\hat{H}t}, ~{\rm for}~ \hat{H}\in {\rm GUE} \big\rbrace\,.
$
%\eeqa
Remarkably, we note that the $k=1$ frame potential with respect to the GUE is given by 
\beqa
 \mathcal{F} ^{(1)}_{\rm GUE}(t) =
 \left\la \!\!\left\la\left\vert\, \tr\, \left(e^{it\hat{H}_\tau}e^{-it\hat{H}_0}\right)\, \right\vert^2\right\ra\!\!\right\ra\, .
\eeqa
This is directly related to the expression of the GUE-averaged Loschmidt echo in Eq. (\ref{L_annealed}) at infinite temperature  as  
\beqa \label{LE_frame}
\la \mathcal{L} (t)\ra=\frac{1}{N^2}\, \mathcal{F} ^{(1)}_{\rm GUE}(t)\quad  {\rm for } \quad \beta=0\, .
\eeqa
This correspondence thus shows that whenever $\mathcal{L}(t) \approx 1/N^2\, $,  the frame potential $\mathcal{F} ^{(1)}_{\rm GUE}(t)\approx 1$, indicating that the GUE time evolution achieves Haar randomness. 
At very early times  $g(0, 0) = N^2$, and the dominant term $g(0,t)^2$  yields $\mathcal{L}(t) \approx 1$ and $\mathcal{F} ^{(1)}_{\rm GUE}(t)\approx N^2$,  showing that the GUE time evolution  is far from achieving Haar randomness. At very late times,  $g(0,t \to \infty) \approx N$, which implies that $\mathcal{L}(t) \approx 2/N^2$ and $\mathcal{F} ^{(1)}_{\rm GUE}(t)\approx 2$. Between these two limits, there is a timescale for which the spectral form factor is about $\sqrt{N}$, yielding  $\mathcal{L}(t) \approx 1/N^2$ as far as  $\mathcal{F} ^{(1)}_{\rm GUE}(t)\approx N^2/(N^2 -1) \approx 1$, and signaling a GUE-mediated Haar scrambling. The relation between $\la \mathcal{L} (t)\ra$ and  $ \mathcal{F} ^{(1)}_{\rm GUE}(t) $ can be extended to other correlation functions, as discussed in  Appendix \ref{AppC}.

However, the relation shown in Eq. (\ref{LE_frame}) is restricted to the infinite temperature limit, since the first frame potential for GUE Hamiltonians  \cite{Cotler2017a} 
\begin{equation}
\mathcal{F}_{{\rm GUE}}^{(1)}(\beta, t) = \frac{1}{N^2{-}1} \Big( g(\beta/2,t)^2 + N^2 - 2 g(\beta/2,t) \Big)\,,
\end{equation}
cannot be directly related to the expression for $\mathcal{L}(t)$ in Eq. (\ref{L_g}) at finite temperature $\beta>0$.

\section{Summary and Discussion}

The work associated with  driving a quantum mechanical system  can be determined via two-projective energy measurements and is
characterized by a  probability distribution, reflecting the stochastic nature of this quantity.
Analyzing the work statistics of complex systems is generally challenging due to  the number of transitions generated during the driving between the energy levels of the initial and final Hamiltonians of the system. Indeed, the latter blows up exponentially as the  size of the system is increased.

In this manuscript, we have reported the exact work statistics of quantum chaotic systems of arbitrary  size that are described by random matrix Hamiltonians.
In particular, we have characterized the mean work and work fluctuations resulting from driving a canonical thermal state by a sudden quench. We  have further shown that the generating function of the work pdf can be interpreted as the Loschmidt echo of the purified thermal state, the thermofield double state. The dynamics of this Loschmidt echo accounts for the reversibility of quenches between chaotic Hamiltonians. 
This dynamics also describes the scrambling of information of the thermofield double state, which acquires a natural interpretation in AdS/CFT being dual to an eternal black hole. The connection between work statistics and information scrambling has been established by mapping the work pdf to the Loschmidt echo, that can be expressed in terms of  the spectral form factor is related to the frame potential.

Our work  establishes an unexpected and firm connection between different fields, including quantum thermodynamics, the emergence of irreversibility, and information scrambling.
This connection is  amenable to experimental tests in a variety of quantum platforms. A prominent example is offered by nuclear magnetic resonance (NMR) experiments, in which measurements of work statistics \cite{Batalhao2014a}, Loschmidt echo dynamics in chaotic systems \cite{Jalabert2001a} and information scrambling \cite{Li2017a} have been reported. Alternative setups include trapped ions \cite{An2015a,Cerisola2017a,Garttner2017a} and superconducting qubits \cite{Georgescu2014a}, where digital quantum simulation techniques \cite{Lanyon2011a,Salathe2015a,Barends2016a} can be used to explore quantum chaos and  systems described by AdS/CFT that exhibit scrambling \cite{Garcia-Alvarez2017a}. 
Chaotic systems could also be implemented in controllable many-body systems by applying random pulses to create an evolution with a Haar-uniform distribution  \cite{Banchi2017a}.

Furthermore, our results should find broad applications in quantum thermodynamics of many-body chaotic systems. 
Quantum many-body systems can be used as a working medium in thermodynamic devices including heat engines and refrigerators \cite{Zheng2015a,Jaramillo2016a,Bengtsson2018a}
and it is natural to explore the role of quantum chaos in this context.

\begin{acknowledgments} 
The authors are indebted to Zhenyu Xu for numerous  discussions that helped to improve the manuscript.
Further, it is a pleasure to thank Luis Pedro Garc\'ia-Pintos for comments on the manuscript. 
We acknowledge funding support from the U.S. Department of Energy, the Spanish Ministerio de Econom\'{i}a y Competitividad (project FIS2015-69512-R), Programa de Excelencia de la Fundaci\'{o}n S\'{e}neca Regi\'{o}n de Murcia (project 19882/GERM/15), the John Templeton Foundation,  and UMass Boston (project P20150000029279). A.C. and A.d.C. acknowledge the hospitality of the Simons Center for Geometry and Physics during completion of this work.
\end{acknowledgments}

\appendix

\onecolumn\newpage

\section{Loschmidt echo averaged over the GUE}\label{AppA}

Starting with the definition of the GUE-averaged Loschmidt echo in Eq. (\ref{L_annealed}), we provide here the details to derive its explicit analytical form, as given in Eq.~(\ref{L_g}). 

The unitary conjugation invariance of the GUE measure allows us to add an  integration  over the unitary group. So denoting by $dV, \,dW$ the Haar measures on the unitary group $\mathcal{U}(N)$,  we follow \cite{Cotler2017a}   and write 
\beqa
\la \mathcal{L} (t)\ra 
 &=& \frac{1}{\la Z(\beta)^2\ra } \int\!\! d\hat{H}_0\, d\hat{H}_\tau\, \rho(\hat{H}_0)\, \rho(\hat{H}_\tau)  \left\vert \tr\, \left(e^{-\sigma_\tau \hat{H}_\tau}e^{-\sigma_0 \hat{H}_0}\right) \right\vert^2\nonumber\\
&=&\frac{1}{\la Z(\beta)^2\ra }\,\int d\hat{H}_0\, d\hat{H}_\tau\, \rho(\hat{H}_0) \rho(\hat{H}_\tau)   \int_{\mathcal{U}(N)}  dV dW
  \Big|\, \tr \big( V^\dagger \Lambda_\tau^\dagger V W^\dagger {\Lambda}_0 W\big)\, \Big|^2\,, 
\eeqa
where  ${\Lambda}_0 \equiv W e^{-\sigma_0 \hat{H}_0 } W^\dagger$  and   $\Lambda_\tau \equiv V e^{- \sigma_\tau \hat{H}_\tau} V^\dagger$ are the matrix exponentials of the GUE initial and final Hamiltonians in the diagonal basis. The first integrals can be expressed in the eigenvalue basis, yielding 
\beqa
& & \la \mathcal{L} (t)\ra=\frac{1}{\la Z(\beta)^2\ra }\, \int D\lambda_\tau \, D\lambda_0    \int_{\rm Haar} dV \, \tr \big( V^\dagger \Lambda_\tau ^\dagger V {\Lambda}_0\big)\,  
 \tr \big( {\Lambda}_0^\dagger V^\dagger \Lambda_\tau  V \big)\,, \nonumber
\eeqa
where we have used the GUE measure defined in Eq. (\ref{GUE_avg}). 
Here, the Haar integrals have been simplified to a single integral using the left and right invariance of the Haar measure. The latter can be explicitly solved by using an exact formula for integrals of monomials of Haar random unitaries \cite{Cotler2017a,Collins2003a, Collins2006a}. Namely, following \cite{Cotler2017a}, we first explicitly write the traces as 
\beqa\label{indices}
\la \mathcal{L} (t)\ra =\frac{1}{\la Z(\beta)^2\ra }\, \int D\lambda_\tau\, D\lambda_0\,  \int_{\rm Haar}  dV 
\Big(V^\dagger_{k_1,k'_1} \Lambda^\dagger_{\tau; k'_1, l_1} V_{l_1, l'_1} \Lambda_{0; l'_1, k_1}  V_{k_2,k'_2} \Lambda^\dagger_{0; k'_2, l_2} V^\dagger_{l_2,l'_2} \Lambda_{\tau; l'_2, k_2} 
\Big)\, ,
\eeqa
and evaluate the Haar integral using the second moment  \cite{Collins2003a, Cotler2017a}
\beqa\label{Haar2}
\int_{\rm Haar} dV \,  V_{l_1, l'_1}  V_{k_2,k'_2}  V^\dagger_{k_1,k'_1}  V^\dagger_{l_2,l'_2} &=&     
{\rm Wg}(1^2,N) \delta_{l_1, k'_1} \delta_{l'_1, k_1} \delta_{k_2, l'_2} \delta_{k'_2, l_2} 
+ {\rm Wg}(1^2,N)  \delta_{l_1, l'_2} \delta_{l'_1, l_2} \delta_{k_2, k'_1} \delta_{k'_2, k_1} \nonumber \\
&+&  {\rm Wg}(2,N)  \delta_{l_1, k'_1} \delta_{l'_1, l_2} \delta_{k_2, l'_2} \delta_{k'_2, k_1} 
+  {\rm Wg}(2,N) \delta_{l_1, l'_2} \delta_{l'_1,k_1} \delta_{k_2, k'_1} \delta_{k'_2,l_2}\, , 
\eeqa
where   we have used the Weingarten functions \cite{Weingarten1978a} given by Collins \cite{Collins2003a} as ${\rm Wg}(1^2,N) = 1/(N^2 - 1)$ and ${\rm Wg}(2,N) =-1/(N(N^2-1))$. 

This simplifies Eq. (\ref{indices}) to 
\begin{eqnarray}
\la \mathcal{L} (t)\ra=\frac{1}{\la Z(\beta)^2\ra }\, \int D\lambda_\tau\, D\lambda_0\, 
\frac{1}{N^2-1} \bigg[&& \tr (\Lambda_\tau^\dagger)\, \tr(\Lambda_\tau)\, \tr( {\Lambda}_0^\dagger)\,  \tr ({\Lambda}_0)  + \tr({\Lambda}_0^\dagger {\Lambda}_0) \tr({\Lambda}_\tau^\dagger {\Lambda}_\tau)  \\
&& - \frac{1}{N} \tr({\Lambda}_0^\dagger {\Lambda}_0)\, \tr (\Lambda_\tau^\dagger)\, \tr(\Lambda_\tau)  
- \frac{1}{N}  \, \tr( {\Lambda}_0^\dagger)\, \tr( {\Lambda}_0)  \tr({\Lambda}_\tau^\dagger {\Lambda}_\tau)\bigg]\, . \nonumber 
\end{eqnarray}
By noting that the integrals in the eigenstate basis can be expressed in terms of the partition function, 
\begin{eqnarray}
 \int D\lambda_i\, \tr \left( {\Lambda_i}^{\dagger} {\Lambda_i} \right) &=&   \Big\langle Z\big(\sigma^{R}_i\big)\Big\rangle\, , \\
  \int D\lambda_i\, \tr ({\Lambda_i}^{\dagger}) \, \tr ({\Lambda_i}) &=&  \Big\langle Z(\sigma_i)Z(\sigma_i^*)\Big\rangle\, , \label{eq:ZZ}
\end{eqnarray}
we obtain the following expression 
\begin{eqnarray} \label{LE_ZZ}
\la \mathcal{L} (t)\ra =\frac{1}{\la Z(\beta)^2\ra }\, \frac{1}{N^2{-}1} 
 \bigg[&&  \Big\langle \! Z(\sigma_\tau)Z(\sigma_\tau^*)\Big\rangle \,  \Big\langle \!Z(\sigma_0)Z(\sigma_0^*)\Big\rangle 
 +\Big\langle Z\big(2 \sigma_0^{R} \big)\Big\rangle \, \Big\langle Z\big(2 \sigma_\tau^{R} \big)\Big\rangle   \\
& & -\frac{1}{N}\Big\langle Z\big(2 \sigma_0^{R} \big)\Big\rangle  \Big\langle Z(\sigma_\tau)Z(\sigma_\tau^*)\Big\rangle 
 -\frac{1}{N} \Big\langle Z(\sigma_0)Z(\sigma_0^*)\Big\rangle  \Big\langle Z\big(2 \sigma_\tau^{R} \big)\Big\rangle  
 \bigg]\, , \nonumber
\end{eqnarray}
where $\sigma^{R}$ denotes the real part ${\rm Re}(\sigma)$. 
This expression can be conveniently written in terms of the spectral form factor as shown in Eq. (\ref{L_g}) of the main text. 

\emph{Alternative derivation.---} The average of the Loschmidt echo over the GUE can be obtained via an alternative route. As any Hermitian Hamiltonian $\hat{H}$ can be diagonalized by a unitary similarity transformation,  it is possible to express the GUE average in Eq. (\ref{GUE_avg}) as an average over the eigenvalues  $\be$ and $N(N-1)$ angles $\btheta$ associated with the unitary \cite{Lobejko2017a}, 
\begin{eqnarray}
\la \hat{O} \ra &=& \int d\hat{H} \, \rho(\hat{H})\, \hat{O} = \int d\be d\btheta \rho_{\be}(\be) \rho_{\btheta}(\btheta)  \, \hat{O}(\be, \btheta)\, ,
\end{eqnarray}
where $\rho_{\be}(\be)$ and $ \rho_{\btheta}(\btheta)$ are   the corresponding distributions given by
\beqa
\rho_{\be}(\be)&=&\frac{2^{N(N-1)/2}}{\pi^{N/2}\prod_{n=1}^Nn!}\prod_{k>j}(E_k-E_j)^2e^{-\sum_kE_k^2}\, ,\nonumber\\
\rho_{\btheta}(\btheta)&=&\frac{1}{(2\pi)^{N(N-1)}}\,.
\eeqa

Since the energy eigenvalues $\be$ and the angles $\btheta$ are statistically independent from each other, the double average in Eq. (\ref{L_annealed}) factorizes as 
\begin{eqnarray} \label{L_alternative}
 \la \mathcal{L}(t) \ra =  \frac{1}{\la Z(\beta)^2\ra }\,\sum_{n,n'}\sum_{m,m'} \left \la p_{m|n}^\tau p_{m'|n'}^\tau \right \ra_{\btheta^0, \btheta^\tau} 
 \left \la e^{-\sigma_\tau E_m^\tau-\sigma_\tau^*E_{m'}^\tau} \right \ra_{\be^\tau} \!
 \left \la e^{-\sigma_0E_n^0-\sigma_0^*E_{n'}^0} \right \ra_{\be^0}\, , 
\end{eqnarray}
where we have made use of the transition probabilities, as in Eq. (\ref{gfguea}). 
While the average of the transition probabilities generates a uniform distribution, known as $\left \la p_{m|n}^\tau \right \ra_{\btheta^0, \btheta^\tau} = 1/N$  \cite{Lobejko2017a}, the average of two transition probabilities is quite more involved and has not yet been reported. 
The difficulty in evaluating this expression comes from the fact that the second transition is not independent from the first one, creating a complex combinational problem. Inspired by the approach in the previous derivation, we here find its explicit form using integration over the unitary group $\mathcal{U}(N)$.

Let us first note the explicit the expression, dropping the indices on the eigenvalues  for clarity, 
\begin{eqnarray}
  \left \la p_{m|n}^\tau p_{m'|n'}^\tau \right \ra_{\btheta^0, \btheta^\tau}   
&=& \Big \la \! \bra{m} U(\tau) \ket{ n}   \bra{n} U^\dagger (\tau)\ket{ m}  
 \bra{m'} U(\tau) \ket{ n'} \bra{n'} U^\dagger(\tau) \ket{ m'}   \! \Big \ra_{\btheta^0, \btheta^\tau}  \nonumber \\
&=& \left \la U_{m,n}(\tau) U^\dagger_{n,m} (\tau)  U_{m',n'}(\tau) U^\dagger_{n',m'}(\tau) \right \ra_{\btheta^0, \btheta^\tau}\,  . 
\end{eqnarray}
Because the GUE measure is invariant under unitary conjugation, we can equivalently evaluate this average over angles as an integral over the unitary group $\mathcal{U}(N)$, i.e. 
\begin{equation}
\int_{\mathcal{U}(N)} dU U_{m,n} U^\dagger_{n,m}   U_{m',n'} U^\dagger_{n',m'}\, .
\end{equation} 
Recognizing here the  second moment of the Haar integral, the evaluation  of this expression can be done using the  Weingarten functions as detailed above and yields
\begin{eqnarray}
\left \la p_{m|n}^\tau p_{m'|n'}^\tau \right \ra_{\btheta^0, \btheta^\tau} =\frac{1}{N^2 -1} \Big( 1 + \delta_{n n'} \delta_{m m'} 
 -\frac{1}{N} \delta_{n n'} - \frac{1}{N} \delta_{m m' } \Big)\, . 
\end{eqnarray}
 This expression can be used to compute the exact average Loschmidt echo  Eq. (\ref{L_g}), upon substituting it into Eq. (\ref{L_alternative}) and evaluating the energy averages.

\section{Spectral form factor}\label{AppB}
We derive here an explicit form of the spectral form factor that provides an alternative form to the result in \cite{delCampo2017a}, with a corrected sign and expression. Starting with the definition of the connected form factor in  (\ref{gConnected}), we expand the square value in Eq. (\ref{eq.TwoLevelCorrelationHermite})  and write 
\begin{eqnarray}\label{gc_app1}
g_c(\sigma) &=& - \sum_{n=0}^{N-1} \sum_{m=0}^{N-1}\int dE dE' \varphi_n(E) \varphi_m(E) e^{-\sigma E}   \varphi_m(E') \varphi_n(E') e^{-\sigma^* E'}   \nonumber \\
&=&  - \sum_{n,m=0}^{N-1}\frac{\mathcal{I}_{n,m}(\sigma)}{2^n n! \sqrt{\pi}} \frac{\mathcal{I}_{m,n}(\sigma^*)}{2^m m! \sqrt{\pi} }\, , 
\end{eqnarray}
where the second line has been obtained using the definition of $\varphi_n(E)$ given in the main text and introducing the integral over the Hermite polynomials 
\begin{equation}
\mathcal{I}_{n,m}(\sigma) = \int dE e^{-E^2 - \sigma E} \mathcal{H}_n(E)  \mathcal{H}_m(E)\, . 
\end{equation}
 Note the symmetrical role played by $(n, m)$ in  this function. 
To find an explicit form, we follow \cite{delCampo2017a} and, for $n \leq m$,  use the identity $\int_\infty dE e^{-E^2}  \mathcal{H}_n(E)  f(E) = \int_\infty dE e^{-E^2} D^n f(E)$ for $f(E) = e^{-\sigma E} \mathcal{H}_m(E)$ to write 
\begin{equation}
\mathcal{I}_{n,m}(\sigma) = \int dE e^{-E^2} D^n( e^{-\sigma E} \mathcal{H}_m(E))\, .  
\end{equation}
The differential $D^n = (d/dE)^n$ can be expanded in a binomial form, $D^n( e^{-\sigma E} \mathcal{H}_m(E)) = \sum_{k=0}^n \binom{n}{k} D^{k} (e^{- \sigma E} )D^{n-k} (\mathcal{H}_m(E))$,  and leads, using the derivatives of the Hermite polynomials $D^{j} (\mathcal{H}_m(E)) = 2^j m! / (m-j)!  \mathcal{H}_{m-j}(E)$, to  a  sum of single-Hermite integral 
\begin{eqnarray} \label{Inm}
\mathcal{I}_{n,m}(\sigma) =  \sum_{k=0}^n  \binom{n}{k} \frac{m!}{(m-n+k)!} 2^{n-k}  (-\sigma)^k 
\int dE e^{-E^2 - \sigma E}  \mathcal{H}_{m-n+k}(E)\, .  
\end{eqnarray}
Using the same identity with $f(E) = e^{-\sigma E}$, the last integral can be evaluated as \cite{delCampo2017a}
\begin{eqnarray}
 \int dE e^{-E^2 - \sigma E}  \mathcal{H}_{m-n+k}(E) = \mathcal{I}_{m-n+k}(\sigma)  
 = \sqrt{\pi} (-\sigma) ^{m-n+k} e^{\sigma^2/4}\, . 
\end{eqnarray}
We further note the simplification of the sum 
\begin{eqnarray}
\sum_{k=0}^n  \binom{n}{k} \frac{x^k}{(m-n+k)!}  &=& \frac{n!}{m!} \sum_{k=0}^n  \binom{m}{n-k} \frac{x^k}{k!}=    \frac{n!}{m!} L_n^{m-n}(-x)
\end{eqnarray}
in terms of the generalized Laguerre polynomials. 
 After symmetrization and defining $\tilde{n}\equiv {\rm min}(n,m)$, this gives Eq. (\ref{Inm}) as 
\begin{equation}
\mathcal{I}_{n,m}(\sigma) =  \sqrt{\pi} e^{\frac{\sigma^2}{4}} (-\sigma)^{|m-n|}\, 2^{\tilde{n}} \, \tilde{n}! \, L_{\tilde{n}}^{|m-n|}(-\sigma^2/2)\, .
\end{equation}
Using this expression and the fact that $\mathcal{I}_{n,m}(\sigma) \mathcal{I}_{m,n}(\sigma^*) = |\mathcal{I}_{n,m}(\sigma)|^2$, we obtain an analytical expression for the connected form factor, Eq. (\ref{gc_app1}),  given in Eq. (\ref{eq.connectedFormFactorGUE}) of the main text.

\section{Loschmidt echo and correlation functions}\label{AppC}

Finally, following \cite{Roberts2017a} we briefly comment on a connection between the GUE-averaged Loschmidt echo and correlation functions at infinite temperature. We focus on a particular time-ordered $2k$-point correlation function
\begin{equation}
\label{OTO_correlator}
 {\rm Tr}\left\lbrace \hat{A}_{1} \, \hat{B}_1(t) \cdots \hat{A}_{k} \, \hat{B}_k(t)\, \hat{\rho} \right\rbrace \,  ,
\end{equation}
where $\hat{A}_1,\cdots ,\hat{A}_k$ and $\hat{B}_1,\cdots ,\hat{B}_k$ are Pauli operators acting on a system of $N_p$ qubits and the dimension of the Hilbert space of the system is $N = 2^{N_p}$. This can be understood as a correlator evaluated at infinite temperature ($\beta=0$), i.e, for the completely mixed state $\hat{\rho}= \mathbb{1} / N$.
Here  $\hat{B}_j(t) \equiv \hat{U}^{\dagger}(t)\, \hat{B}_j\, \hat{U}(t)$, with the unitary $\hat{U}(t) \in \mathcal{E}_t^{\rm GUE}$. 
The average of these kind of out-of-time ordered correlation functions for an ensemble of unitary operators such as $\mathcal{E}_t^{GUE}$  %$\left\langle \langle \hat{A}_{1} \, \hat{B}
can be written as
\begin{eqnarray}
\label{OTO_averaged}
\int_{\mathcal{E}_t^{GUE}}\, d\hat{U}\, {\rm Tr}\left\lbrace  \hat{A}_{1} \, \hat{B}_1(t) \cdots \hat{A}_{k} \, \hat{B}_k(t)\, \hat{\rho}\right\rbrace \, .
\end{eqnarray}
These ensemble averaged correlators were shown to be related to the  $k-$th frame potential through \cite{Roberts2017a}
\begin{flalign} \label{OTO_frame}
\frac{N^{4k}}{N^{2(k+1)}}\, \mathcal{F}^{(k)}_{\rm GUE} =
\sum_{\left\lbrace\hat{A}\right\rbrace\, \left\lbrace \hat{B}\right\rbrace }\left|  {\int_{\mathcal{E}_t^{GUE}}} d\hat{U}\, {\rm Tr}\left\lbrace  \hat{A}_{1} \, \hat{B}_1(t) \cdots \hat{A}_{k} \, \hat{B}_k(t)\, \hat{\rho}\right\rbrace  \right|^{2}\,  , 
\end{flalign}
where the sum is over the $N^{4k}$ possible Pauli operators $\hat{A}_1,\cdots ,\hat{A}_k\, ,\hat{B}_1,\cdots ,\hat{B}_k$. Thus, considering the case for $k=1$   
and the correspondence established in Eq. (\ref{LE_frame}), it is possible to express the GUE-averaged Loschmidt echo in terms of an averaged 2-point function
\begin{equation}
\left\langle  \mathcal{L}(t)\right\rangle =\frac{1}{N^2}\!\!\! \sum_{\left\lbrace\hat{A}\right\rbrace\, \left\lbrace \hat{B}\right\rbrace }\left| \int_{\mathcal{E}_t^{GUE}}\, d\hat{U}\, {\rm Tr}\left\lbrace  \hat{A}\, \hat{B}(t)\, \hat{\rho} \right\rbrace  \right|^{2}\, ,
\end{equation}
both measuring through their decrease, the degree of randomness induced by GUE-time evolution.

%%%%%%%
%\nocite{apsrev41Control}
%\bibliography{main} % References file
%\bibliographystyle{apsrev4-1}
%\bibliographystyle{plainnat}


\end{document}